\documentclass{article}

\newif\ifpdf
   \ifx\pdfoutput\undefined
   \pdffalse
\else
   \pdfoutput=1
   \pdftrue
\fi

\ifpdf

\usepackage{ae}

\usepackage[pdftex]{graphicx}

\else

\usepackage[T1]{fontenc}

\usepackage[dvips]{graphicx}

\fi

\usepackage{arxiv}

\usepackage[T1]{fontenc}    
\usepackage{hyperref}       
\usepackage{url}            
\usepackage{booktabs}       
\usepackage{amsfonts}       
\usepackage{nicefrac}       
\usepackage{microtype}      
\usepackage{lipsum}

\usepackage{latexsym}
\usepackage{psfrag}
\usepackage{amsmath}
\usepackage{amssymb}
\usepackage{amsthm}
\usepackage{xypic}
\usepackage{graphicx}
\usepackage{tabularx}
\usepackage{longtable}
\usepackage[active]{srcltx}
\usepackage{gensymb}

\usepackage{subfigure}

\newcommand{\htwho}{HTaWO$_6$.nH$_2$O}

\newcommand{\ltw}{LiTaWO$_6$}

\newcommand{\cm}{cm$^{-1}$}

\newcommand{\wo}{WO$_6$}
\newcommand{\tao}{TaO$_6$}

\newcommand{\etal}{et.al.}

\title{The {\it H$_x$Li$_{1-x}$TaWO$_6$.nH$_2$O} trirutile\\ structure characterization through Raman and \\IR spectra comparison: \\
The non-center symmetric case}

\author{
  D.~Valim$^{1}$,
  AG Souza Filho,$^{2}$ J M Filho,$^{2}$ \\
  $^{1}$Faculdade de Ci\^encias Exatas e Tecnologia,\\
  Universidade do Estado de Mato Grosso,\\
  $^{2}$Departmento de F\'isica, Universidade Federal do Cear\'a, \\
  \texttt{$^\ast$daniel.valim@unemat.br} \\
   \And
 OL Alves,$^{3}$, MAC de Santis,$^{3}$,  E N Silva$^{4}$ \\
  $^{3}$Departamento de Qu\'imica-UNICAMP/Campinas, Brazil\\
  $^{4}$Departamento de F\'isica, Universidade Federal do Maranh\~ao, S\~ao Lu\'is, Brazil\\
   \\
}

\begin{document}
\maketitle

\begin{abstract}
We have performed the LiTaWO$_6$ vibrational studies for both Raman and infrared actives modes. Although we do not have crystalline samples, the study was conducted qualitatively, firstly comparing the sample with similar compounds in the literature, and secondly comparing both Raman and infrared spectra of the LiTaWO$_6$ compound. Although the procedure is limited, we can assign 8 of the 10 Raman-IR active modes expected by Group Theory.
\end{abstract}

\keywords{Trirutile \and Non-center Symmetric \and Raman actives modes\and IR actives modes}

\section{Introduction}

We can understand the trirutile structure like a normal rutile superlattice\cite{Bolzan}. The standart composition is $A_{1/3}B_{2/3}O_{2}$ or $AB_{2}O_{6}$ where cations $A^{n+}$ and $B'^{n+}$ where in under the condition of $n+2n=12$ equation. The possible combinations for $n$ and $n'$ are $n=6$, $n'=3$ and $n=2$, $n'=5$. For this cases, the difference of three charge units($n'-n=3$) is enough to ionic ordering three times the normal rutile c axis. The trirutile Space Group is the same as simple rutile $P4_2/mnm$($D_{4h}^{14}$), $Z = 2$. In this Group A and B cations occupy $2a(000)$ and $4e(00z)$ Wyckoff positions, respectively, while there two non-equivalent Wyckoff positions for 4f(xx0) and 8j(xxz) of oxygens. A distorted trirutile structure (P$2_1/n$), whose is a P$4_2/mnm$ subgroup, was found for SnO$_2$,  CrTa$_2$O$_6$ \cite{Martinez2009,Arevalo-Lopez2008,Maddox2006,Jiang2004} and CuSb$_2$O$_6$ compounds\cite{Maimone2018}.

The LiMWO$_{6}$ ($M=Ta, W$) compound belongs to a family material with chemical formula like AB$^{'}$B$^{"}$O$_6$, where A, B$^{'}$ and B$^{"}$ are monovalent, three and hexavalent, respectively. In 1970, Blasse and Paw\cite{Blasse1970} proposed the these materials belong to tetragonal center-symetric structure to LiTaWO$_6$ P$42_1m$ ($D_{2d}^3$) (with $a = b = 4.6776, c=9.0337$ \AA). Fourquet \etal\cite{fourquet} proposes the non-center symmetric Space Group P$\bar{4}2_1$m (D$_{2d}^3$) (with a = b = 4.6776, c = 9.2710 \AA) for LiTaWO$_6$ compound. More recently, in 2002, Catti\cite{catti2002} proposed a new structure to this compound, obtaining an orthorhombic center-symmetric C$mmm$ ($D_{2h}^{19}$) (with a = b = 6.6088 and c = 9.2999 \AA) with volume approximately two times the ones before proposed. In this work the HTaWO$_6$.H$_2$O trirutile  was refined to P$4_{2}/mnm$ ($D_{4h}^{14}$) Space Group. In all cases, the LiTaWO$_6$.H$_2$O trirutile is constituted by LiO$_6$, TaO$_6$ and WO$_6$ octahedras connected like showed in Figure\,\ref{LABEL}. Note that due to the octahedra ordering, Li, Ta and W cations are disposed parallel to c axis.

As a matter of fact, the P$4_2/mnm$, P$\bar{4}2_1m$ and C$mmm$ Space Groups obey the well defined Group-subgroup mathematical relationships, when a reduction of symmetry of P$42_1m$ or C$mmm$ must be spliting of Wyckoff positions as showed in Table\,\ref{tab:diagr1}.

The introduction of two similar size cations in B site positions can result in ordering of structure as occur to LiTaWO$_6$ trirutile. It means that taking in account the Space Group correlation between P$4_2/mnm$ and P$\bar{4}2_1m$ there are two non-equivalent positions to B cations and three non-equivalent positions to oxigen anions.

It is known that HTaWO$_6$ crystallizes in two different structures, a defect pyrochlore and trirutile. The defect pyrochlore one can be obtained via ionic change of its alcaline compound counterpart. In this case it is the KTaWO$_6$ compound. Instead, the HTaWO$_6$ trirutile is formed by anion change reaction from LiTaWO$_6$ of same structure.

Buvanesh et. al. \cite{Bhuvanesh} has been studied non-linear optic responses of LiM$^{V}$M$^{VI}$O$_6$ (M$^{V}$=Nb, Ta; M$^{VI}$=Mo, W) policrystaline samples with 25-45$\,\mu m$ grain sizes. It was demonstrated that materials exibhit SHG (Second Harmonic Generation) of 1064$\,nm$ wavelength radiation with efficiency of 16-28 times bigger than $\alpha-$quartz. We know that, SHG is allowed only to non-center-symmetric samples what is consistent with Space Group described above (P$\bar{4}2_1m$). 

\section{Experimental Characterization}
\label{sec:headings}

\subsection{Synthesis}
The reactants used to prepare LiTaWO$_6$ precursor samples are shown in Table\,\ref{tab:first}. The synthesis of LiTaWO$_6$ Trirutile precursor was based in the procedure described by Catti\cite{catti1993} using stoichiometric reaction with following reactants:

\begin{table}[ht]
    \centering
    \begin{tabular}{|l|c|c|c|}
    \hline \hline 
    Reactant & Formula & Manufacturer & Purity(\%) \\
    \hline
     Tantalum Dioxide   & Ta$_2$O$_5$ & Alfa & 99  \\
     Tungsten Dioxide   & WO$_3$ & Aldrich &  99  \\
     Lithium Carbonate   &  Li$_2$CO$_3$ & Vetec  & 99  \\
     Nitric Acid    &   HNO$_3$ &   Merck   &   65*  \\
    \hline
    \end{tabular} 
    \\ *Aquous soluction concentration.
    \caption{Reactants used to trirutile samples synthesis.}
    \label{tab:first}
\end{table}

, with following reaction describe bellow:

\begin{equation*}
   Li_2CO_3 +Ta_2O_5 + 2WO_3 \xrightarrow{} 2LiTaWO_6 +CO_2    
\end{equation*}

The reactants were crushed and mixed in an agate mortar for 10 minutes, and after the mixing was transferred for a Platine crucible to heating at 850\,$\celsius$ in an oven. The system remains under heating by 24 hours, with an interruption after 12 hours for crushing and homogenization of the mixing. 

The LiTaWO$_6$ Trirutile precursor was submitted at ionic exchange reactions in HNO$_3$ 4mol.L$^{-1}$ solution for 48 h, at 80\,$\celsius$. The supernatant solution was exchanged in an analogous way describe in reference\cite{santis} to Pyrochlore synthesis. The used acids and experimental parameters are based on procedures described in the literature\cite{mari1986a,catti1993}. We have performed Li$^+$/H$^+$ ionic exchange reactions in LiTaWO$_6$ precursor to yield the HTaWO$_6$ trirutile. 

Finally, the protonated compounds were again subjected to ion exchange reactions, this time the proton was replaced by Li$^+$ ions. The reactions were performed using the procedures of LQES laboratory for similar compounds\cite{GARCIA-MARTIN,zhui1996}. For this time, the reactions were performed by only one Li:H molar ration of 1:5 to form H$_ x$Li$_{1-x}$TaWO$_6$.  We know the ion exchange reaction, to 1:5 for example, occurs as follows

\begin{equation*}
5HTaWO_{6} + 1LiTaWO_{6} \xrightarrow{} 6H_{0.833}Li_{0.167}TaWO_6    
\end{equation*}

Where we can adjust the x, which is the concentration ratio of lithium to 0.167, wich result H$_{0.833}$Li$_{0.167}$TaWO$_6$.nH$_2$O trirutile compound. This procedure is analogous to that described in the references\cite{Valim2009,Valim2018}.

\subsection{X-ray diffraction}

The X-ray diffractogram were obtained in a Shimadzu diffractometer, operating in scan mode with CuK$\alpha$ radiation, generated at 40 KV and 30 mA current. The scanning speed used was 2/min in 2$\theta$, with accumulation for reading every  0.6 seconds. The slits were used: divergent 1.0 mm and collection 0.3 mm. The calibration of the scanning angle was done with polycrystalline silicon and the samples were analyzed in powder form.

\begin{figure}[htp]
 \begin{center}
  \centering
   \includegraphics[scale=0.37]{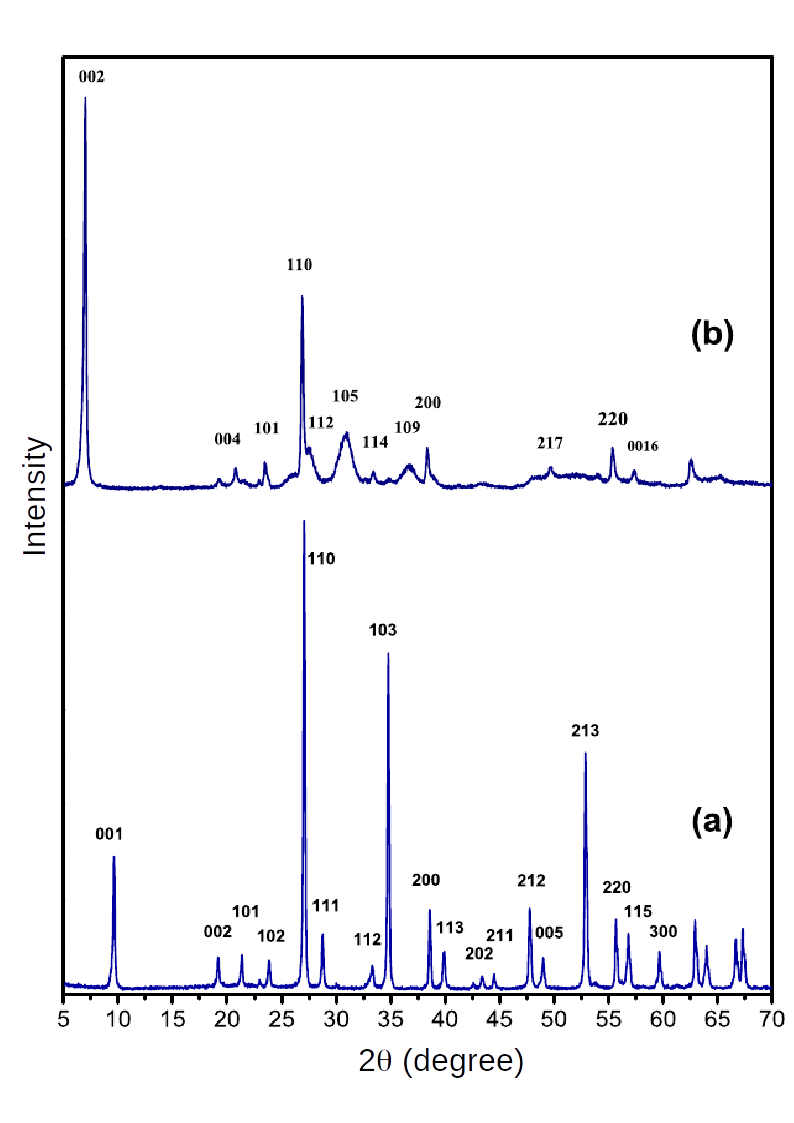}
\caption{\small{X-ray diffractograms of compounds in the trirutile structure P$\bar{4}$2$_1$m (D$_{2d}^3$): (a)\,\ltw\, (b)\,\htwho\,\cite{santis}.}}
\label{fig:raios-x}
 \end{center}
\end{figure}

\begin{figure}[ht]
    \centering
    \mbox{
    \subfigure[100 direction]{\label{LABEL-A}
    \includegraphics[scale=0.4]{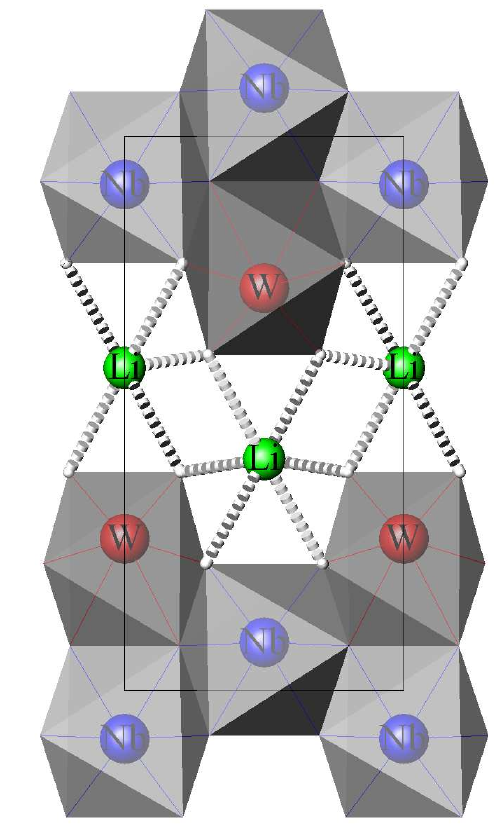}
    }
    \subfigure[001 direction]{\label{LABEL-B}
    \includegraphics[scale=0.25]{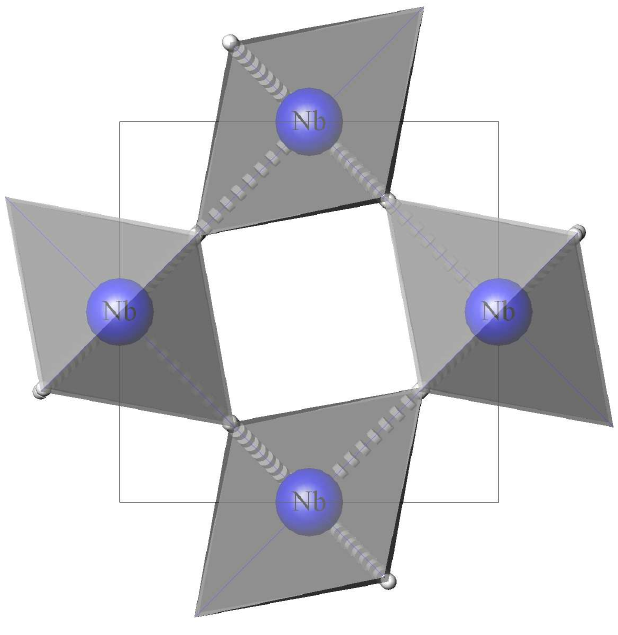}}
    }
    \caption{Unit cell projections of P$\bar{4}$2$_1$m (D$_{2d}^3$) Space Group of  LiNbWO$_6$ compound\,\cite{fourquet} at long (a) [100] and (b) [001] directions.\label{LABEL}}
\end{figure}

\subsection{FTIR spectroscopy}

The Fourier Transform Infrared (FTIR) spectra were obtained  by Bomen ABB-FTLA 2000 in the 250-4000\,\cm interval with 4\,\cm \,resolution and 16 accumulations. The samples were studied in KBr pellets from Fluorolube dispersion, for spectra in the 1300-4000\,\cm, and in Nujol, for 250-1300\,\cm, using cesium-iodide windows. The FTIR measurements were made in the LQES laboratory\cite{santis}.

\subsection{Raman spectroscopy}

All spectra measurements had been performed at room temperature through following experimental apparatus: an Argon laser, a Jobin Yvon T64000 three-monochromator spectometer equipped by a microscope with 180$\,mm$ of focal objective. The laser line was 488$\,nm$ with power of 300$\,mW$ on the trirutile samples.

\section{Results and Discussions}
\subsection{X-ray diffraction}

\paragraph{}The ionic exchange does not change the structure of the compound, like is showed in the diffractograms of \textbf{Figure \ref{fig:raios-x}} wich strutures correspond that showed in the Figure\,\ref{LABEL}. The ionic exchange Li$^+$/H$^+$ in the trirutile structure, the tetragonal crystalline system produce the expansion of the cell c parameter from 9.24 to 25.52\,\AA, as we can verify by Miller index $00l$. Instead of, we got a significant dislocation of $002$ peak position, while the $110$ peak position remains unaltered.

For a lot of authors, the c parameter expasion is the primary evidence of tetragonal cell remaining after ionic exchange reactions\cite{bhat,kumada,kumada1053}. The formation of supercell can be attributed to interlayers translations of water molecules, perpendicular to the c axis, that were oriented to optimize hydrogen bonds connected to adjacent layers\cite{kumada}. A thorough analysis of the characteristics of that reactions has shown that Li$^+$/H$^+$ exchange reactions process occur together with a tetragonal primitive (LiTaWO$_6$) to tetragonal body-center(HTaWO$_6$.H$_2$O) transformation\cite{bhat}.

In the \textbf{Figure \ref{fig:raios-x}} are showed the X-ray diffractogram of both trirutile pristines HTaWO$_6$.nH$_2$O and LiTaWO$_6$ of the reference\,\cite{santis}. The ionic change Li$^+$/H$^+$ in the trirutile structure produce the cell expansion in the parameter c, from 9.27\,\AA\, to 25.62\,\AA\, like verified by Miller index shift $001$. Inspite of remarkable shift of $002$ peak, the position of $110$ diffraction peak remains unchanged. For the majority of authors, it is the main evidence of tetragonal cell keeping after ionic change\cite{bhat,kumada,kumada1053}.

The formation of superlattice is assign at translation of interlamellar layers, perpendicular at c-axis, because introduction of water molecules into structure\,\cite{bhat,kumada}. The ionic change Li$^+$/H$^+$ in compound, also cause alteration of primitive tetragonal unit cell to body centered tetragonal, when HTaWO$_6$.nH$_2$O is its hidrated form ($x= 3/2>n>1/2$)\cite{bhat}. The structure maintenance after Li$^+$/H$^+$ ionic change, reveals high bi-dimensional mobility of precursor lithium, confirmed by studies of its difusion coeficient\,\cite{Sebastian}.

\begin{longtable}[c]{ccc} 
\caption{\small{Outspread of Wyckoff positions from  
trirutile \\structures (D$_{4h}^{14}$) to similars (D$_{2d}^{3}$) and (D$_{2h}^{19}$).}\label{tab:diagr1}} \\ \hline
    \small\xymatrix@-1.8pc{
    AB^{'}B^{''}O_6 & & AB_2O_6 & & AB^{'}B^{''}O_6\\
    P\bar{4}2_1m (D_{2d}^3)& &\ar[ll]P4_2/mnm (D_{4h}^{14})\ar[rr]& & Cmmm (D_{2h}^{19})\\
     & &     & & A(2a)\\
    A(2c)& &A(2a)\ar[urr]\ar[drr]\ar[ll]& &      \\
     &  &    & & A(2c)\\
    B^{'}(2c)& &     & & B^{'}(4k)\\
      & &B(4e)\ar[ull]\ar[dll]\ar[urr]\ar[drr] & &      \\
    B^{''}2(2c)& & &      & B^{''}(4l)\\
     &  &  &   & O1(4h)\\
    O1(4e)& &O1(4f)\ar[urr]\ar[drr]\ar[ll] & &      \\
     & &    &  & O2(4i)\\
    O2(4e)& &    &  & O3(8n)\\
      & &O2(8j)\ar[ull]\ar[dll]\ar[urr]\ar[drr]& &      \\
    O3(4e)&  &   &  & O4(8o)\\
}\\ \hline
\end{longtable}

\section{Group Theory and vibractional spectroscopy}
\label{sec:others}
In the case of ABBO$_6$ compounds, the Table\,\ref{tab:tgd2d} shows 54 degrees of freedom through D$_{2d}$ representation Factor Group of P$\bar{4}2_1m$ Space Group (Z = 2). From all modes ($9A_1+3A_2+3B_1+9B_2+15E$) the $9A_1+3B_1+8B_2+14E$ are Raman active modes, while the $8B_2+14E$ are infrared active modes and $B_2+E$ correspond at acoustic modes. In contrast the $3A_2$ is a remainded silent mode.

\begin{table}[ht]
\centering\caption{Group Theory analysis of D$_{2d}$ factor group to AB$^{'}$B$^{''}$O$_6$ compounds.} \label{tab:tgd2d}
\begin{tabular}{ccc}
\hline\hline
ions &Wickoff sites/Symmetry & Irreducible Representations \\
\hline
A& 2c/C$_{2v}$ & A$_1\oplus$B$_2\oplus$2E  \\
B$^{'}$& 2c/C$_{2v}$ & A$_1\oplus$B$_2\oplus$2E  \\
B$^{''}$& 2c/C$_{2v}$ & A$_1\oplus$B$_2\oplus$2E  \\
O& 4e/C$_{s}$  & 2A$_1\oplus$A$_2\oplus$B$_1\oplus$2B$_2\oplus$3E \\
O& 4e/C$_{s}$  & 2A$_1\oplus$A$_2\oplus$B$_1\oplus$2B$_2\oplus$3E \\
O& 4e/C$_{s}$  & 2A$_1\oplus$A$_2\oplus$B$_1\oplus$2B$_2\oplus$3E \\
\hline
\multicolumn{3}{l}{$\Gamma_T$=9A$_1\oplus$3A$_2\oplus$3B$_1\oplus$9B$_2\oplus$15E} \\
\multicolumn{3}{l}{$\Gamma_R$=9A$_1\oplus$3B$_1\oplus$8B$_2\oplus$14E~~~~~~~~~~~~~~~~$\Gamma_{IR}$=8B$_2\oplus$14E} \\
\multicolumn{3}{l}{$\Gamma_s$=3A$_2$~~~~~~~~~~~~~~~~~~~~~~~~~~~~~~~~~~~~~~$\Gamma_{ac}$=B$_2\oplus$E}\\
\hline\hline
\end{tabular}
\end{table}

The drawback in this methodology is its use in only polarized light. Since our samples are ceramics or powder, in consequence without crystal axis orientation, it is impossible to perform polarized measurements in order to discuss the vibrational modes in terms of irreductible factor groups. The more apropriated method is analysis in and out modes into structures. Both pyrochlore and trirutile structures are composed of octahedron networks BO$_6$ (and BO$_6$ or B/B'O$_6$) and cations A (and B or B/B').

In general the vibration modes of octahedron are observed between 300-900\,cm$^{-1}$ whereas extern mode frequencies are observed below 300\,cm$^{-1}$. In this case we should make a correlation table symmetry for lattice, sub-lattice and lattice itself, like showed in \textbf{Table\,\ref{tab:diagr1}}. However, we should know what and how much internal modes are allowed in a free octahedron for subsequent analyses by Group Theory based on symmetry reduction of octahedron inside lattice by correlation of crystal structures in study. The Group Theory calculation of vibrational modes of molecules type XY$_6$ is showed in Apendixe B of Reference \cite{eder2008}. The distribution of the 21 degree of freedom of octahedral molecule is given by 
\begin{equation*}\label{eq:gammaoct}
    \Gamma_v = A_{1g}(\nu_1) \oplus E_g(\nu_2) \oplus
3F_{1u}(T,\nu_3,\nu_4) \oplus F_{1g}(L) \oplus F_{2g}
(\nu_5)\oplus F_{2u}(\nu_6)
\end{equation*}where $\nu_1$, $\nu_2$ e $\nu_3$ modes are streching moves of B$^{''}$ bonds, and $\nu_4$, $\nu_5$ and $\nu_6$ angular deformations O - B$^{''}$ - O. The libration octahedra modes L(F$_{1g}$) and $\nu_6$ that originally are silent in both Raman and IR spectra can be observed due lowing local symmetry octahedra when put in a crystalline lattice. How is the LiTaWO$_6$ case considering its P$\overline{4}2_1m$ Space Group, like shown in the  \textbf{Table\,\ref{tab:diagr2}}. The B$^{''}$O$_ 6$ that originally was in the cubic symmetry O$_h$ is in the C$^{'}_{2v}$ now. In consequence, it make L(F$_{1g}$) and $\nu_6$ active in both Raman and IR spectra. 

This methodology has a limitation that all B$^{''}$O$_6$ octahedra to be un-conected therselves, although all B$^{'}$O$_6$ are connected.  The vibrational activity for boths octahedra depend of bond forces B$^{'}$ - O and
B$^{''}$ - O.  In this case, the B$^{''}$ cation has more eletrovalence and make the B$^{''}$ - O constant force stronger than B$^{'}$ - O. In the case of both constants forces are approximately close therselves, we can observe both octahedral activities.

Catti et. al.\cite{catti1990} study the thermal, structural, vibrational and electrical properties of HTaWO$_6$.xH$_2$O (x = 3/2, 1/2, 0). The results obtained through X-ray showed that HTaWO$_6$.3/2H$_2$O belongs to tetragonal system with lattice parameters a = 4.710\,\AA\, and a = 25.80\,\AA.

\begin{figure}[htbp]
\centering
\includegraphics[width=2.35 in]{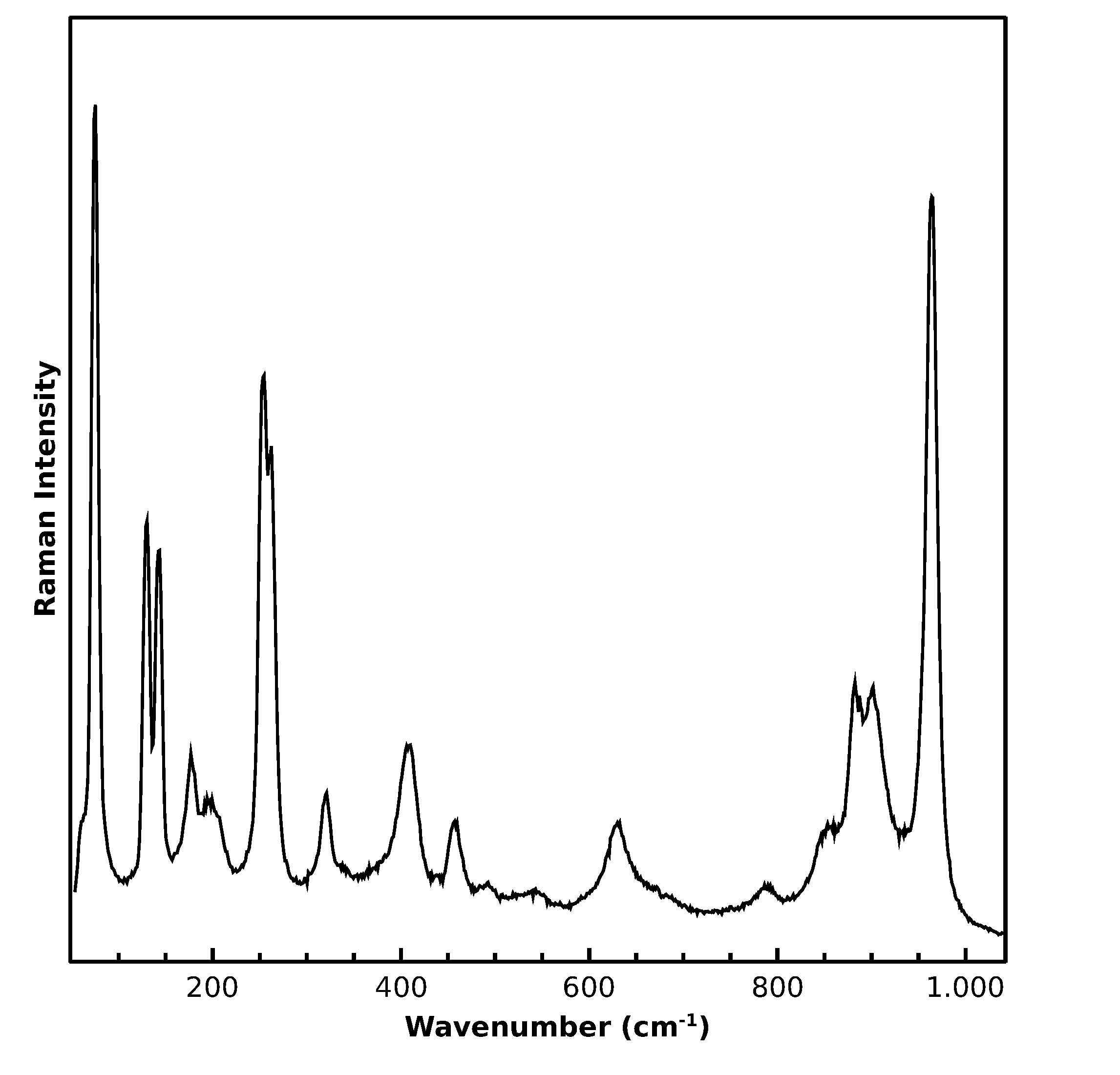}
\caption{LiTaWO$_6$ Raman Spectrum at room temperature.}\label{fig:raman-ltw}
\end{figure}

The \textbf{Figure \ref{fig:raman-ltw}} show LiTaWO$_6$ Raman spectrum at room temperature. This spectrum is in agreement with one reported by Catti \etal\cite{catti1993}, showing a similarity bethween the samples. Inspite that, there is no deep discution about observed vibration modes, due the absence of the LiTaWO$_6$ monocrystals. There is a only reference about 960\,\cm \,band, which has been atributed to TaO$_6$ octahedra stretch vibration that share sides.

Not all predicted modes have been observed in the Raman spectrum, maybe sobreposition of some bands, or due low polarizability of some modes. Thus, the amount of bands and their line widths are consistent with ordered crystalline structure where atoms occupy positions of low symmetry.

\begin{longtable}[l]{cccc}
\caption{\small{Correlation for AB$^{'}$B$^{''}$O$_6$ with D$_{2d}$ point group. Raman (R), Infrared (IR), and Silent modes (S).\label{tab:diagr2}}} \\ \hline
\small\xymatrix@-1.8pc{
     ion/sym.~free~ion& site ~symmetry & cel.~unit.~symm & vibrational-modes~-activity\\
 A,B^{'}/- & cC^{'}_{2v}(2) & \textbf{D}_{2d} & \\
    &(2Tz)A_1\ar[r]\ar[ddddddr]& A_1\ar[r] & 9A_1(3T, \nu_1,\nu_2, \nu_3, \nu_4, \nu_5, \nu_6)-R \\
    &(2Tx)B_1\ar[dddddddr]& & \\
    &(2Ty)B_2\ar[ddddddr]& A_2\ar[r] & 3A_2(L, \nu_5, \nu_6)-S \\
B^{''}O_6/O_h& cC^{'}_{2v}(2)& &\\
    (\nu_1) A_{1g}\ar[r]& A_1\ar[ddr]\ar[uuuur]&  B_1\ar[r] & 3B_1(L, \nu_5, \nu_6)-R \\
    (\nu_2) E_g\ar[dddr]\ar[ur]& & & \\
    (R) F_{1g}\ar[r]\ar[ddr]\ar[ddddr]& A_2\ar[uuuur]\ar[uur]&B_2\ar[r] & 9B_2(2T, \nu_1,\nu_2, \nu_3, \nu_4, \nu_5, \nu_6, ac)-R,IR\\
    (T, \nu_3, \nu_4)3F_{1u}\ar[dr]\ar[dddr]\ar[uuur]&  &&\\
    (\nu_5)F_{2g}\ar[ddr]\ar[uur]\ar[uuuur]& B_1\ar[r] &E\ar[r] & 15E(5T, 2L, \nu_2, 2\nu_3, 2\nu_4, \nu_5, \nu_6, ac)-R,IR\\
    (\nu_6)F_{2u}\ar[uuur]\ar[ur]\ar[uuuuur]& &&\\
    & B_2\ar[uur]& &}\\ \hline
\end{longtable}

\subsection{FTIR and Raman spectra comparison}

\begin{figure}[ht]
\centering
 \includegraphics[width=4 in]{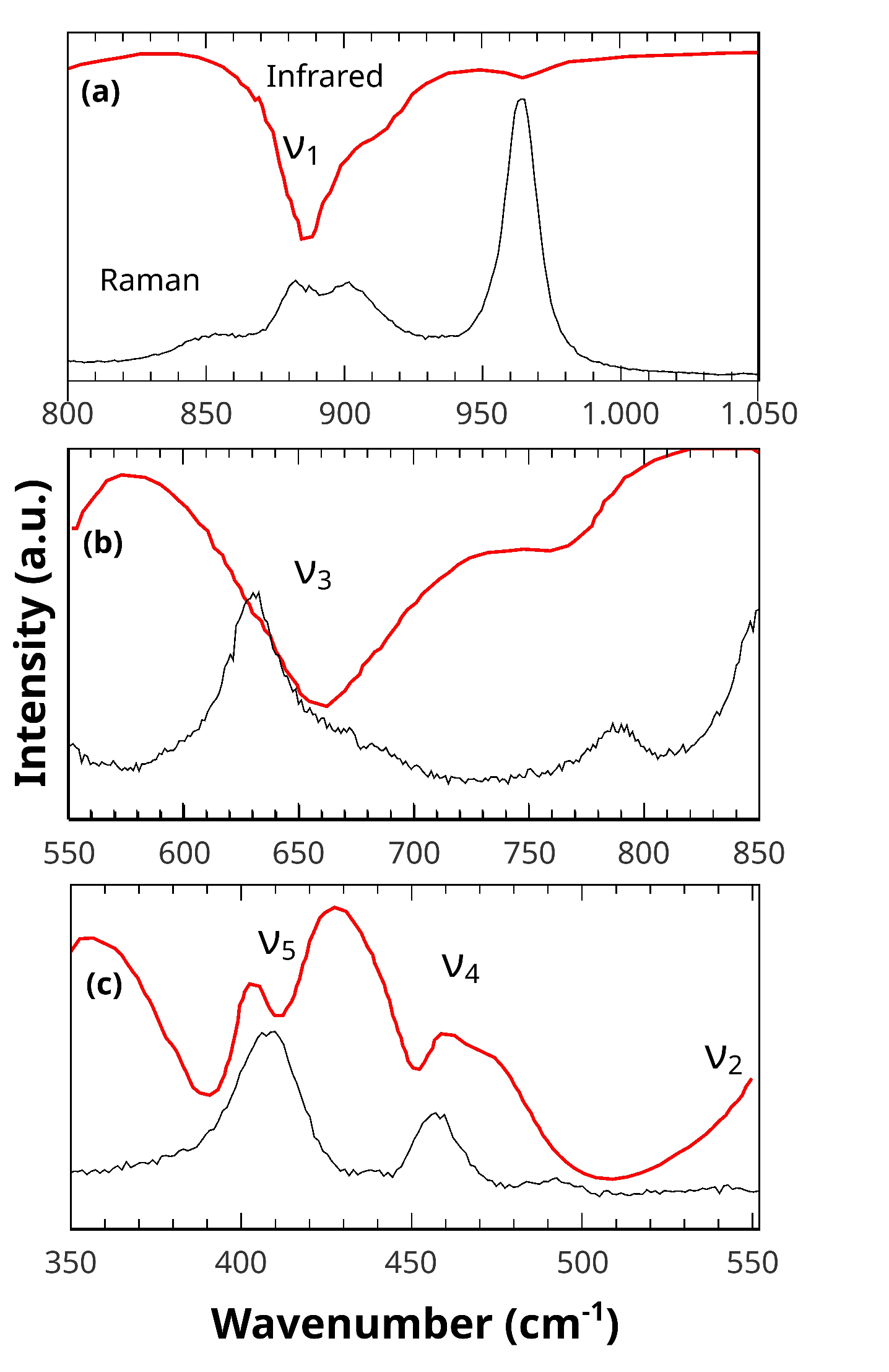}
 \caption{Comparison between Raman and infrared spectra of LiTaWO$_6$ between (a) 800-1050\,cm$^{-1}$, (b) 550-840\, cm$^{-1}$ and (c) 350-550\, cm$^{⁻1}$ ranges.}\label{fig:ltwramanir}
\end{figure}

In the Figure\,\ref{fig:ltwramanir} are compared both IR transmitance(reproduced by Ref\,\cite{santis}) and Raman spectrum of LiTaWO$_6$ measured at room temperature. With respect to infrared spectrum, we can observe ten bands localized on 320, 390, 410, 451, 510, 660, 760, 885, 910, and 966\,\cm. Note that, a lot of this bands have a clear correlation with some Raman spectrum bands, that is expected if we take in account a non-center-symmetric structure. In fact, in according to Bhuvanesh \etal\cite{Bhuvanesh}, this compound exhibit second harmonic generation, an exclusive propertie of polars materials, whose inversion of symmetry center is absense. For this results, the more probable structure for LiTaWO$_6$ is P$\overline{4}2_1m$.

Although we did not have oriented monocrystals of LiTaWO$_6$, we could doing a qualitative analysis of the Raman modes observed based on both octahedra (B$^{''}$O$_6$) and cations lattice vibrations in the structure. Since the W-O bond lenths are aproximately equal to Ta-O bonds, the vibrational spectrum of LiTaWO$_6$ can be dominated mainly by WO$_6$ or TaO$_6$ octahedron vibrations. In some cases, if charges or masses of  B$^{'}$ e B$^{''}$ are very differents, it should be possible make a differentiation between vibration of B$^{'}$O$_6$ and B$^{''}$O$_6$. In general, the frequency order of bonds stretching is   
$\nu_1>\nu_3\geq\nu_2$, while for frequencies of angular deformations O-B$^{''}$-O is $\nu_4\geq\nu_5>\nu_6$. In the \ltw \,structure the \wo \,or (\tao)\, structure are slighly distorted, leaving itselfes to lower symmetry  C$_{2v}$, than pyrochlore one (D$_{3d}$). The influence of octahedral symmetry lowering in the vibrational spectrum should be analyzed using a correlation diagram like is showed in the Table\,\ref{tab:diagr1}. In according with Table\,\ref{tab:diagr1} we observed 21 Raman modes that correspond to octahedra internal modes (2$\nu_1$, 3$\nu_2$, 4$\nu_3$, 4$\nu_4$, 4$\nu_5$,
4$\nu_6$) of which thirteen ($\nu_1$, 2$\nu_2$, 3$\nu_3$, 3$\nu_4$, 2$\nu_5$, 2$\nu_6$) must be observerd for both Raman and infrared, while the remaining only observed by Raman spectroscopy.

In according to literature, for compounds containing tantalum and tungstein, like peroviskite, for example, LiTaO$_3$\,\cite{Kostritskii}, A$_2$InTaO$_6$
(A = Ba, Sr)\cite{Dias2007}, A$_2$CoWO$_6$ (A = Ba, Sr)\cite{Ayala2007} and A$_3$In$_2$WO$_9$ (A = Ba, Sr)
\cite{Silva2009}, the wavenumbers of the symmetrical breath of the octahedrons \tao~ and \wo~ have values of 840\,\cm. Base on it, the unusual characteristic of the \ltw~ spectrum are bands localized between 810 and 1000\,\cm~. Similar characteristic are observed on  TiTa$_2$O$_7$ \cite{eror} and TiNb$_2$O$_7$ \cite{MCCONNELL}, that have two octahedra distribuction of MO$_6$ (M = Ta e Nb), respectively; one with the shared vertex MO$_6$ and other with shared edges (M$_2$O$_{10}$). Two intense bands were observed in 899 and 1020 \cm~for the TiTa$_2$O$_7$ and 1000 and 892 \cm~for the TiNb$_2$O$_7$ compounds. The first was attributed to molecular breath symmetric mode M$_2$O$_{10}$, while the second was designed like a molecular breath symmetric of the M$_2$O$_{11}$ unit. The wavenumber diferences between the two bands was 121 and 180\,\cm to TiTa$_2$O$_7$ and TiNb$_2$O$_7$, respectively. If this wavenumber difference is systematic for the octahedra shared vertex and edge, the corresponding Raman bands are 964 and 851\,\cm (See Figs. \ref{fig:raman-ltw} and \ref{fig:ltwramanir}), which difference is 113\,\cm. It may be a indicative that TaO$_6$ and WO$_6$ are well correlated. There is also a doublet to assign; it is centered in 890\,\cm. Since the crystalline system of octahedra \tao~ and \wo~ in the trirutile structure is different of the peroviskites, we expected a displacement of the wavenumber of the vibrations to these octahedra, what it implies the 890\,\cm~doublet may be a symmetric streching ($\nu_1$) of the isolated octahedra \tao~and \wo.

In the next spectral region, showed in the \textbf{Figure\,\ref{fig:ltwramanir}(b)}, we can observe two bands in 600 and 760\,\cm in the infrared spectrum and three bands in the Raman spectrum in 630, 670 and 785\,\cm. In general, in this region is localized the assymetric stretching of infrared active mode $\nu_3$. Lavat and Baran \cite{Lavat} studied infrared spectra of several perovskites containing tantalum as manly octahedral unit and observed that this mode is localized on 650 and 670\,\cm~ region. On the other hand, Liegeois-Duyckaerts and
Tarte\cite{LIEGEOIS} studied the Raman and infrared spectrum of the other series of perovskites containing tungstein as mainly octahedral unit and observed that $\nu_3$(WO$_6$) occured in the 600-650\,\cm~ region. Based on this, we can only assign the bands between 650\,\cm~ to $\nu_3$(WO$_6$)+$\nu_3$(TaO$_6$). Despite of suppositions, we can not assign the 750-810\,\cm~ bands based on the free octahedra, but we can suppose this vibrations have origin in movements of couple (TaWO$_{11}$ or by (TaWO$_{10}$) octahedra.

The \textbf{Figure \ref{fig:ltwramanir}(c)} show the Raman and Infrared spectra comparison in the 350-570\,\cm~ region. In this spectral region are concentrated stretching anti-symmetric modes of M - O ($\nu_2$) bonds, and angular deformations of O-M-O angles, with symmetry of $\nu_4$ and $\nu_5$ modes of the free octahedron. In according to literature, the $\nu_4$(TaO$_6$)+$\nu_4$(WO$_6$), $\nu_5$(TaO$_6$)+$\nu_5$(WO$_6$)
e $\nu_2$(TaO$_6$) modes have energy in the regions of 330-450\,\cm~ \cite{Lavat,LIEGEOIS}, 375-450\,\cm \\\cite{Dias2007,LIEGEOIS} and 550\,\cm \cite{Dias2007}. We observed five and six bands in infrared and Raman spectra, respectively. In general, the $\nu_2$ mode have low intensity \cite{Dias2007} and in lot of cases is not observed \cite{LIEGEOIS} in Raman spectrum, however we can assign the 537\,\cm~ band in the \ltw~ Raman spectrum as $\nu_2$(TaO$_6$) symmetric mode. The wide band centering in 510\,\cm~ is very comom in a lot of compounds of peroviskites family \cite{Silva2009,Lavat} having $\nu_4$ symmetry mode, thus we assign the bands between 430-520\,\cm as a $\nu_4$ symmetry modes. Finally the bands localized between 375-425\,\cm~are vibrations of $\nu_5$ symmetry.

The bands lower than 375\,\cm are the most difficult to indentify because in this region are localized the librational ($L$) and translational ($T$) modes of crystal lattice. In this case, we consider the movement of the cations Li, Ta and W at long $x$, $y$ or $z$ directions, as showed in the Table \ref{tab:diagr1}. Through the results of the Group Theory shown in this papper we must observe sixteen modes (10T+3L+3$\nu_6$) in this specrtral region. Due the degenerescence of some bands or the weak intesity of the Raman spectrum due a material polarizability, we only observe ten bands in this region.

\section{Raman spectrum of the H$_{0.833}$Li$_{0.167}$TaWO$_6$ trirutile}

\begin{figure}[ht]
\centering
 \includegraphics[width=4 in]{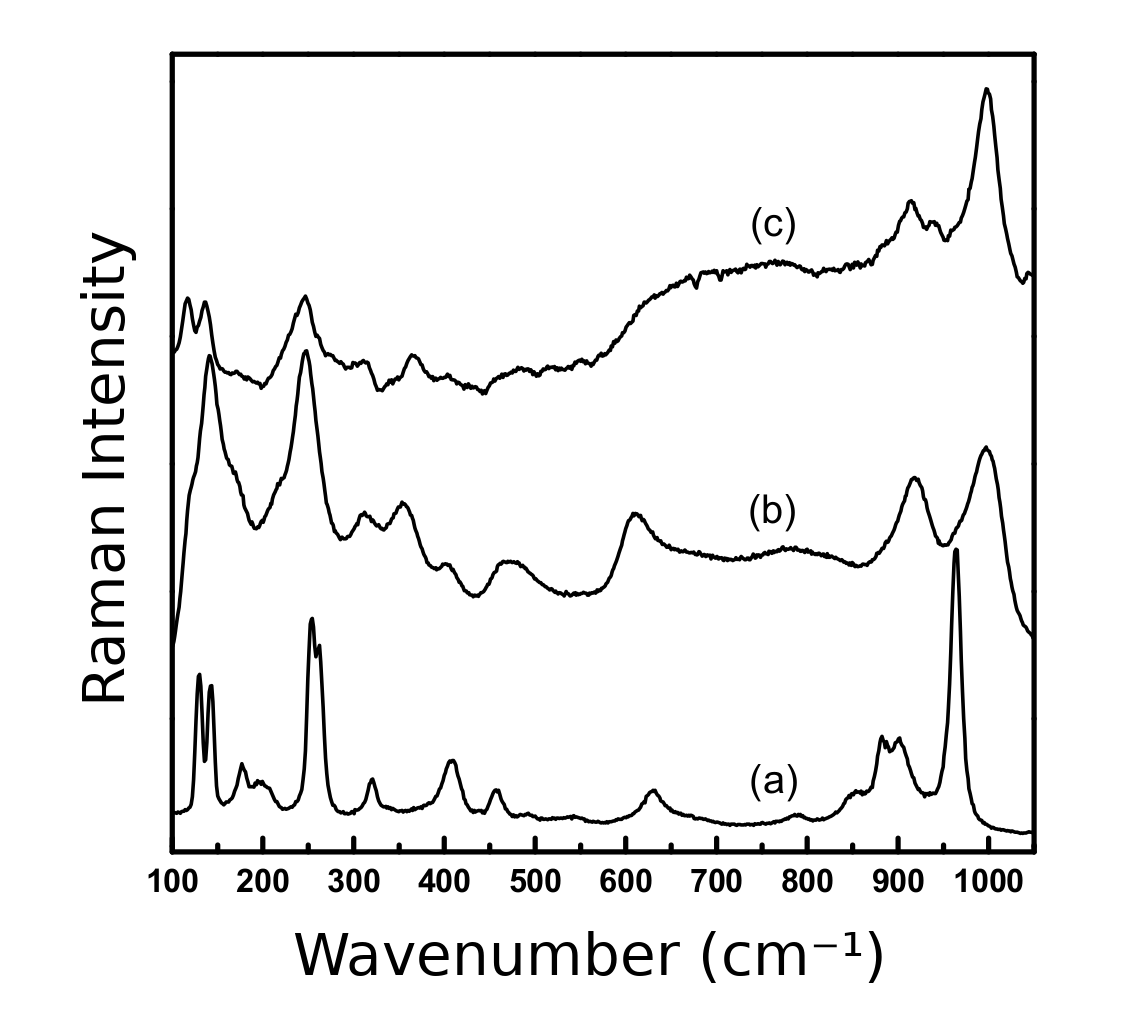}
\caption{Raman spectra of the  trirutile compounds: (a)  LiTaWO$_6$ sample, (b) HTaWO$_6$.nH$_2$O [35] and (c) doped sample H$_{0.833}$Li$_{0.167}$TaWO$_6$.}\label{trirutiles}
\end{figure}

\textbf{Figure \ref{trirutiles}} shows the Raman 
Spectra of the H$_{0.833}$Li$_{0.167}$TaWO$_6$, LiTaWO$_6$, 
and \\HTaWO$_6$.nH$_2$O\cite{santis}. The Raman spectra of the H$_{0.833}$Li$_{0.167}$TaWO$_6$ and HTaWO$_6$.nH$_2$O show 
some differences with LiTaWO$_6$ for both the number of bands 
and bandwidth.

The Raman spectrum of \textbf{Figure\,\ref{trirutiles}(b)} is well represented for HTaWO$_6$.nH$_2$O trirutile with n between $3/2$ and $1/2$ like shown in the reference\,\cite{catti1993}. We can see the band shift of higher energy to high wavenumbers.

When hydration of HTaWO$_6$.nH$_2$O increases the structure tend to become more open with c lattice parameter changing from 20 to 26\, \AA\,\cite{bhat,mari1988}. While with the Li$^+$ ion the structure tends to more closely, with c lattice parameter remaining to 9.30\,\AA. The supercell formation also is attributed the translation of the interlayers of the water molecules, perpendicular to the c axis, that are oriented to optimize the hydrogen bonds that connected to adjacent layers\cite{kumada}. The detailed analysis of the characteristic of the ionic exchange reactions had been shown for some authors. They took into account the ionic exchange reactions Li$^+$/H$^+$ occur followed by unit cell change, from tetragonal primitive to body center tetragonal (HTaWO$_6$.nH$_2$O)\cite{bhat}. According to Kumada et.al.\cite{kumada} the Li$_{0.9}$H$_{0.1}$TaWO$_6$ compound, have lattice parameter c=11.1\,\AA \,that is close to the LiTaWO$_6$ due to the large Li$^+$ ion concentration. As H$_{0.833}$Li$_{0.167}$TaWO$_6$ has low Li$^+$ concentration, it is expected that the vibrational spectrum of this compound resembles that of HTaWO$_6$.nH$_2$O, as observed in \textbf{Figure\,\ref{trirutiles}}.

Although the HTaWO$_6$.nH$_2$O have the trirutile structure with c lattice parameter almost three times c lattice parameter of the LiTaWO$_6$ compound, we still see a strong disorder in the octahedral cations Ta$^+$ and W$^+$ due to the enlargement of the high wavenumber bands (more than 500 cm$^{-1}$), taking as reference the LiTaWO$_6$ Raman spectrum.

As the H$^+$ and Li$^+$ ions contribute only to lattice libration modes (less than 500 cm$^{-1}$) in the wavenumber interval showed in \textbf{Figure\,\ref{trirutiles}}, we observe a remarkable difference in this spectrum region.

\section{Conclusion}

We can summarize that studies that indicate a non-symmetric P$\bar{4}2_1m$ structure can be performed, although done in a polycrystalline sample. Even so, we conducted a qualitative study using the knowledge about Raman modes of TiTa$_2$O$_7$ and TiNb$_2$O$_7$ by comparison. 

The prediction by Group Theory for both IR and Raman active modes are 13 ($\nu_1$, 2$\nu_2$, 3$\nu_3$, 3$\nu_4$, 2$\nu_5$, 2$\nu_6$) from the total of 21 internal Raman modes  (2$\nu_1$, 3$\nu_2$, 4$\nu_3$, 4$\nu_4$, 4$\nu_5$,
4$\nu_6$).   With respect to the infrared spectrum,
we observed ten bands localized on 320, 390, 410, 451, 510, 660, 760, 885, 910, and 966\,\cm, with a lot of these bands, have a clear correlation with some Raman spectrum bands, that is expected to a non-center-symmetric structure, as described by  Bhuvanesh \etal\cite{Bhuvanesh}. 

As long as there is no significant difference between B'O$_6$ and B"O$_6$ mass and charge, the band frequencies of B'O$_6$ and B"O$_6$ are closely each other, like a consequence of both B'-O and B"-O are approximately equal. It is a case of Ta-O and W-O bonds. As a consequence, the vibrational spectrum of LiTaWO$_6$ is dominated mainly by WO$_6$ and TaO$_6$ octahedron vibrations. Take into account the frequency order of bonds stretching is $\nu_1>\nu_3\geq\nu_2$, while for frequency order of angular deformations O-B$^{''}$-O is $\nu_4\geq\nu_5>\nu_6$, we can assign:

\begin{itemize}
    \item $\nu_1$ to 851 and  964\,\cm;
    \item The 510\,\cm band belong to $\nu_4$, as well as the bands localized between 430-520\,\cm;
    \item The $\nu_5$ bands are localized between 375-425\,\cm; 
    \item The bands lower than 375\,\cm are more difficult to identify. In this spectral region are localized the librational($L$) and translational ($T$) modes. 
\end{itemize}
We can observe ten infrared bands, (320, 390, 410, 451, 510, 660, 760, 885, 910, and 966\,\cm),  of which 8 we can assign for both Raman and infrared activity.

Although this procedure is very limited, we can assign some vibrational bands for both Raman and infrared activity, with a significant level of confidence to conclude that our sample belongs to P$\bar{4}2_1m$ Space Group.

\section{Acknowledgments}

D. Valim acknowledge E.N. Silva for his conversations about Group Theory and vibrational modes. D. Valim, AG Souza Filho, JM Filho, OL Alves, M.A.C. de Santis, and E.N. Silva acknowledge the Brazilian agencies, CNPq, CAPES, FUNCAP, and FAPESP for financial support.


\end{document}